\documentclass{article}
\usepackage{spconf,amsmath,graphicx}

\def\x{{\mathbf x}}

\title{Quality Assurance Challenges for Machine Learning Software Applications During Software Development Life Cycle Phases}
\name{Md Abdullah Al Alamin and Gias Uddin, DISA Lab, University of Calgary}
\address{}
\def\bf{\textbf}

\def\fig {Figure~}

\def\sec {Section~}

\def\it{\textit}

\def\tt{\mct}

\let\OLDthebibliography\thebibliography
\renewcommand\thebibliography[1]{
  \OLDthebibliography{#1}
  \setlength{\parskip}{0pt}
  \setlength{\itemsep}{0pt plus 0.3ex}
}

\newcommand{\mct}[1]{{\footnotesize {\texttt {#1}}}}

\usepackage{paralist}

\newcommand{\nd}{\vspace{1mm}\noindent}

\begin{document}
%
\maketitle
\begin{abstract}
In the past decades, the revolutionary advances of Machine Learning (ML) have
shown a rapid adoption of ML models into software systems of diverse types. Such
Machine Learning Software
Applications (MLSAs) are gaining importance in our daily lives. As such, the Quality Assurance (QA) of MLSAs is of paramount importance. Several research efforts are dedicated to determining the specific challenges we can face while adopting ML models into software systems. However, we are aware of no research that offered a holistic view of the distribution of those ML quality assurance challenges across the various phases of software development life cycles (SDLC). This paper conducts an in-depth literature review of a large volume of research papers that focused on the quality assurance of ML models. We developed a taxonomy of MLSA quality assurance issues by mapping the various ML adoption challenges across different phases of SDLC. We provide recommendations and research opportunities to improve SDLC practices based on the taxonomy. This mapping can help prioritize quality assurance efforts of MLSAs where the adoption of ML models can be considered crucial. 
\end{abstract}
\begin{keywords}
Machine Learning Software Application (MLSA), SDLC, ML Pipeline, Challenge, Quality Assurance
\end{keywords}

\section{Introduction}

Each disrupting change in
software development required the software industry to evolve and adapt a novel
strategy. The latest trend is the widespread interest in the adoption of ML (Machine Learning) capabilities into large scale Machine Learning software applications (MLSAs) \cite{amershi_se_case_study_2019}. The market of MLSA and Artificial Intelligence (AI) is expected to grow at a compound annual growth rate of 29.7\% 
worldwide.   However, quality assurance challenges for MLSAs are hard to address, leading to deadly errors. For example, recently, a Uber self-driving car ran into pedestrians because the sensor could not detect those. There is now a growing concern on the quality assurance of 
safety-critical ML applications like self-driving
cars~\cite{chen2015deepdriving}, healthcare, financial institutions, etc.

Software quality assurance is a systematic approach to detect defects and it can improve the
reliability and adaptation of MLSA~\cite{zhang_ml_testing_survey_horizons_2020}. However, ML models' stochastic nature and their dependency on data introduce diverse novel challenges like testing non-determinism in the MLSA outputs~\cite{zhang_ml_testing_survey_horizons_2020}. While deep learning (DL) models have revolutionalized ML models' performance across domains, DL models often behave like a black-box, which makes it difficult
to be evaluated and explained in safety-critical
applications~\cite{willers2020safety}. 

In traditional software development, first we gather requirements. We then design, develop, test, deploy and maintain the
application. For ML systems, we
still need to scope out the goal of the application, but instead of designing the
the algorithm we let the ML model learn the desired logic from data~\cite{amershi_se_case_study_2019}. Such observations lead to the question of whether and how ML models can be adopted without disrupting the software development life cycle (SDLC) of the MLSAs. Ideally, \textit{ML workflow/pipeline} and \textit{SDLC} phases should go hand in hand to ensure proper quality assurance. However, as we noted above, such expectations can be unrealistic due to the inherent differences in how ML models are designed and how traditional software applications are developed. We thus need a holistic view of the diverse ML adoption challenges we encounter during the SDLC phases of an MLSA development. Such insights can be used to improve the ML adoption pipeline and the SDLC phases of MLSA development. 

In recent years, significant research efforts are devoted to understanding the diverse quality assurance challenges while adopting ML models into software systems~\cite{devanbu2020deep, santhanam2019engineering, feldt2018ways}. However, we are not aware of any research that specifically focused on mapping the QA challenges across SDLC phases. In this paper, by conducting an in-depth literature review of the challenges of ML adoption and by determining how such challenges may permeate, we have produced a taxonomy of quality assurance challenges that practitioners face during the adoption of ML models into the diverse SDLC phases. We present the taxonomy and describe it with examples. We conclude by offering recommendations for research opportunities that lie ahead to ensure the quality assurance of MLSAs.

\section{Background and Related Work} \label{sec:background}
We introduce two concepts (SDLC and ML pipeline stages) on which this paper builds on. We then briefly discuss related work. More related work are discussed in \sec\ref{sec:MLSA_challenges}. 

\begin{figure*}
\centering
\vspace{-10mm}
\includegraphics[scale=.60]{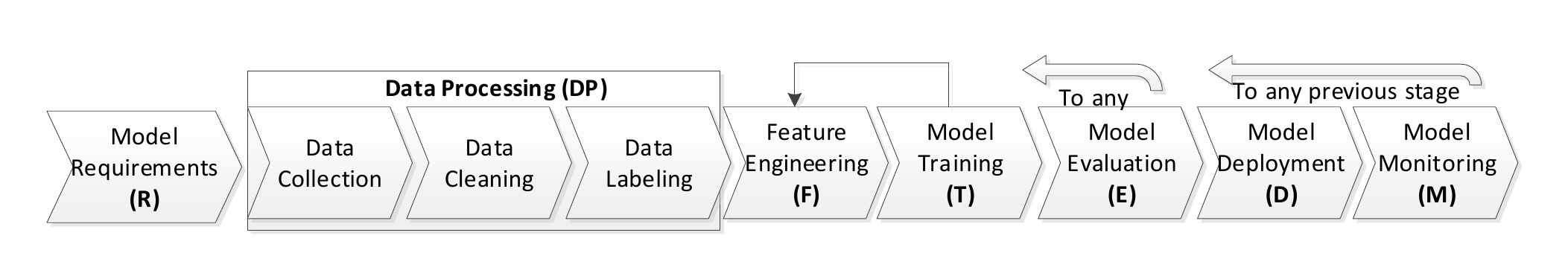}
\vspace{-5mm}
\caption{ A typical iterative pipeline (i.e., workflow) to prepare an machine learning (ML) model for a system \cite{amershi_se_case_study_2019} }
\label{fig:ml-pipeline}
\vspace{-5mm}
\end{figure*}

\nd\bf{$\bullet$ Background.} SDLC methodology is embedded in traditional application development that consists of Requirement analysis, Designing \& planning, Implementation, Quality assurance, Deployment, Maintenance phases. Our interest is to study the QA challenges faced during different SDLC phases of MLSA. Machine learning workflow is a little bit different from traditional application development. It begins with system requirement analysis to data collection and processing, feature engineering and training, evaluation of model performance and then model deployment and monitoring. Figure \ref{fig:ml-pipeline} shows a typical ML model development stages/workflows \cite{amershi_se_case_study_2019}. It is an iterative process based on model's performance.

\nd\bf{$\bullet$ Related Work.} There are quite a few numbers of research that focus on the current challenges of machine learning in software engineering\cite{devanbu2020deep, feldt2018ways, santhanam2019engineering}, empirical studies on best practices of integrating AI capabilities\cite{amershi_se_case_study_2019}, developers survey on challenges faced during different SDLC\cite{ml_practice_2019}. There are also quite a few research on the quality assurance of ML models like reliability, transparency, trustworthiness, etc. \cite{saria_safe_reliable_2019, roscher2020explainable, goodfellow2014explaining, santhanam2019engineering, roscher2020explainable, qiu2019review}. Zhang et al.\cite{zhang_ml_testing_survey_horizons_2020} provides a comprehensive survey on ML testing considering 138 related papers on four aspects of MLSA testing such as what properties to test, what ML components to test, testing workflows and test scenario of different type of ML application. Shafiq et al.\cite{shafiq2020machine} conducts a bibliometric study and provides an MLSA taxonomy of the adaptation of ML techniques across different SDLC phase. It focuses on the research attention at different SDLC of MLSA. Lack of modularity in ML models compared to traditional software can make those hard to debug~\cite{zhang_ml_testing_survey_horizons_2020}. Test oracle problem, due to insufficient specification and data dependency, can introduce a wide range of challenges during software testing \cite{barr2014oracle}. While the above papers offer information of ML adoption challenges, none unlike us, has focused on providing a holistic view of the ML adoption challenges within the standard development life cycle of MLSAs.

\section{Quality Assurance Challenges in MLSA} \label{sec:MLSA_challenges}
\begin{figure*}
\centering
\includegraphics[scale=0.55]{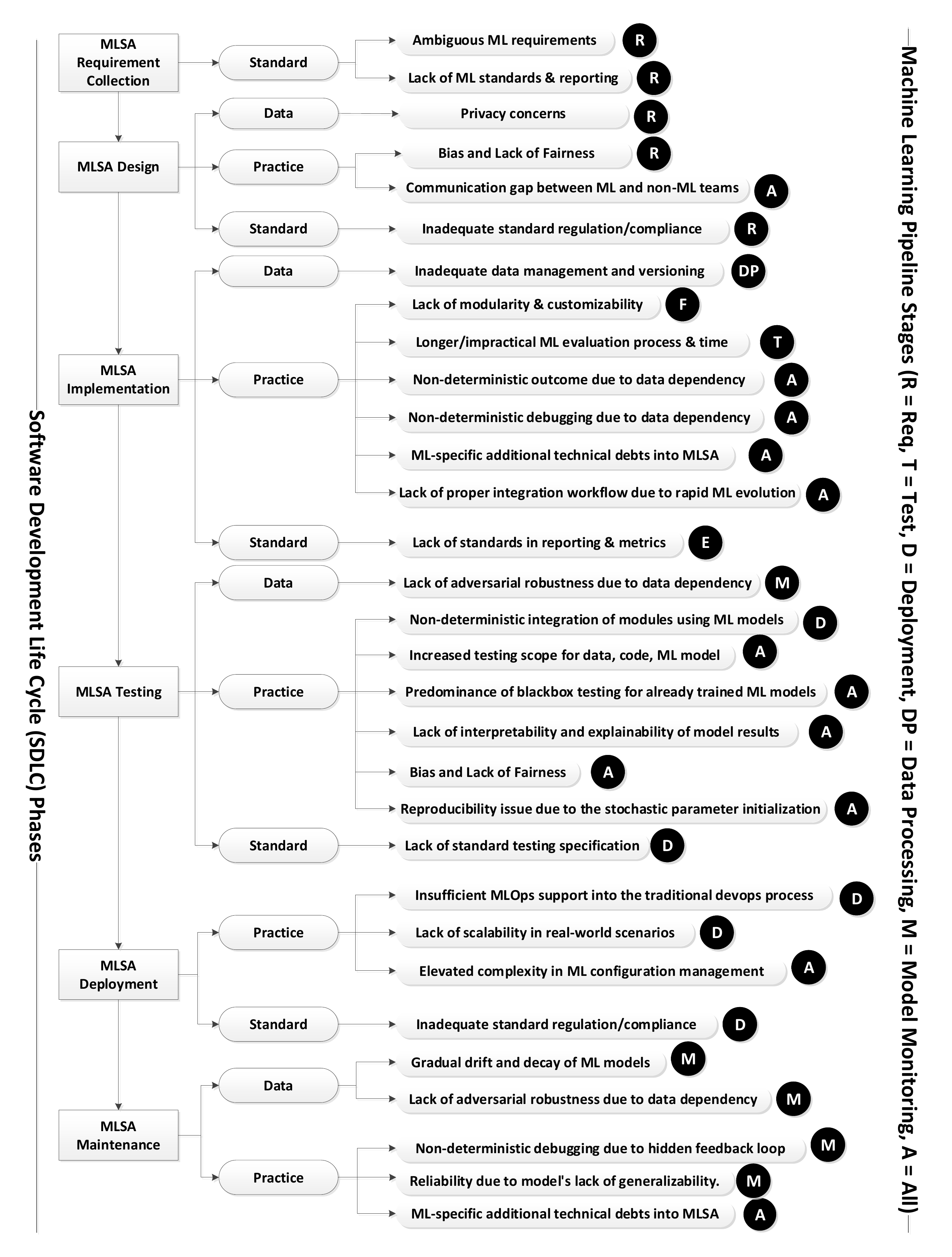}
\caption{A taxonomy of MLSA quality assurance challenges across different phases of SDLC}
\label{fig:taxonomy_MLSA}
\end{figure*}
In \fig\ref{fig:taxonomy_MLSA}, we present a taxonomy of quality assurance challenges for MLSA development. The challenges are derived from an in-depth literature review of quality assurance challenges for ML adoption into software systems. We group the challenges under six SDLC phases: \begin{inparaenum}[(1)]
\item Requirement analysis, 
\item design and planning,
\item implementation, 
\item testing, 
\item deployment, and 
\item maintenance. 
\end{inparaenum} For each challenge, we also show the specific stages in ML pipeline that are impacted/consulted to address the challenge. The ML pipeline stages are derived from Amershi et al.~\cite{amershi_se_case_study_2019}. 

In total, we found 31 challenges that we group into three higher categories: \begin{inparaenum}[(1)]
\item \textit{Data}. It contains QA challenges related to data collection, cleaning, labelling, management,
\item \textit{Practice}. It contains QA challenges that is faced in practice by SE teams.
\item \textit{Standard}. This category of QA challenges represent the challenges due to the lack of Standard specification or guidelines.
\end{inparaenum} A challenge can be observed in multiple SDLC phases. For example, lack of ML standards and metrics can be a problem during MLSA requirement collection as well as during MLSA implementation phase. However, depending on the types of SDLC, the same challenge can be observed in varying formats. For example, while lack of standards in ML model evaluation may result in ambiguities during MLSA requirement collection, it can also affect the development team during their analysis of whether an ML model is good enough with regards to the stated requirements and designs. We now discuss the challenges per SDLC phases below.

\nd\bf{$\bullet$ SDLC Phase 1. MLSA requirement Collection.} 
Two challenges arise due to the lack of clarity in \textit{Standard} specification:  \begin{inparaenum}[(1)]
\item \textit{Ambiguous requirement (ML stage = R, i.e., Requirements)}: MLSAs can have \textit{ambiguous specification}, as it is difficult to define the expected behaviour, which can change due to data~\cite{ml_practice_2019, finkelstein2008fairness}. 
\item \textit{Lack of standards in reporting \& metrics (R)}: MLSAs do not have a standard reporting specification like other industries (transportation, aviation).
\end{inparaenum}

\nd\bf{$\bullet$ SDLC Phase 2. Design and planning.}  There are four challenges under three categories: Data, Practice, and Standard. \textit{Data} category has one challenge. \begin{inparaenum}[(1)]
\item \textit{Privacy concerns (R)}: Many companies can not use raw user-data because of terms and service agreement with the users, organizational legal and ethical constrains. Inability to understand the internal representation of the model might cause privacy breach~\cite{castelvecchi2016can}. 
Federated learning~\cite{mcmahan2017communication} has the potential to deliver these features properly~\cite{li2020federated}.
\end{inparaenum} \textit{Practice} category has two challenges. \begin{inparaenum}[(1)]
\item \textit{Bias and Lack of Fairness (R)}: An MLSA can be biased if the training is biased~\cite{mehrabi2019survey}.
\item \textit{Communication gap between ML and non-ML teams/practitioners (ML Stage = A, i.e., all)}: Qualitative and quantitative studies suggest that it is important to adopt best of SE and ML research world such as sharing data~\cite{ml_practice_2019}, complexity reduction, deletion of features, improving reproducibility~\cite{sculley_hiddent_debt_2015}.
\end{inparaenum} There is one challenge faced due to lack of \textit{Standard} specifications: \textit{Inadequate standard regulation/compliance (R)}: During training we need to consider user data privacy and confidentiality. Organizations need to comply with user data protection acts HIPPA\cite{annas2003hipaa}, GDPR~\cite{voigt2017eu}.

\nd\bf{$\bullet$ SDLC Phase 3. MLSA Implementation.}  There are eight QA challenges. The \textit{Data} category has one challenge: \textit{Inadequate/Inefficient data management and versioning support (for ML stage = DP, i.e., Data Processing)}. There is a lack of tools to manage different data attributes (e.g., data freshness). The \textit{Practice} category has six challenges: \begin{inparaenum}[(1)]
\item \textit{Lack of modularity \& customizability (ML stage = F, i.e., Feature Engineering)}: Due to lack of customizability, it requires a significant effort to reuse a model for a different task or handle different input format. 
\item \textit{Longer/impractical evaluation process \& time of ML models (ML Stage = T, i.e., Testing)}: The performance of a model can not be evaluated until training on the whole dataset is finished, which is computationally expensive and time-consuming. 
\item \textit{Non-deterministic outcome due to data dependency (A)}: ML models are inherently dependent on training data. However, real-world data distribution may be different, and the model might misbehave. \item \textit{Non-deterministic debugging due to data dependency (A)}: The debugging strategy of traditional systems is not quite applicable for many ML-systems: bugs might be in the code as well as in the data. 
\item \textit{Lack of well-defined workflow for integration due to rapid ML evolution (A)}: There is a lack of ML-specific process management tools, or specification to estimate time and resources. 
\item \textit{Presence of additional technical debt due to ML adoption into software (A)}: A mature ML systems can be of 95\% of glue code that mainly connects different library and packages \cite{sculley_hiddent_debt_2015, SE_DL18}.
\end{inparaenum} There is 1 QA challenge under \textit{Standard} category: \textit{Lack of standards in reporting \& metrics (E)}. There is a lack of standards to compare the performance of different models, guideline to identify \& report fault~\cite{dlse_furute_2020}.

\nd\bf{$\bullet$ SDLC Phase 4. MLSA Testing.} We find eight challenges. \textit{Data} category has one challenge: \textit{Lack of adversarial robustness due to data dependency (ML Stage = D, i.e., Deployment)}. Adversarial example refers to a small malicious modification of the input that causes the model to erroneous output~\cite{madry2017towards}. The \textit{Practice} category has six challenges: \begin{inparaenum}[(1)]
\item \textit{Non-deterministic integration of ML models (ML stage = M, i.e., monitoring)}: All the external systems that provide input or consumes the output of the ML model should be explicitly monitored because these dependencies can affect the ability to release updates\cite{SE_DL18, sculley_hiddent_debt_2015}.
\item \textit{Increased testing scope due to the needs for both data and code testing (A)}: QA for ML systems are difficult as they are inherently non-deterministic and they self-learn~\cite{amershi_se_case_study_2019}.
\item \textit{Predominance of black-box-type testing for already trained ML models (A)}: Various phases of a typical MLSA contain highly coupled components. Faults occurred in one component of this pipeline may propagate to other phases and thus hard to detect and fix \cite{sculley2014machine, felderer_qa_overview_challenges_2021}.
\item \textit{Lack of interpretability and explainability of model results (A)} is prevalent when features are too abstract for human/tester to understand~\cite{lipton2018mythos, ribeiro2016should}. 
\item \textit{Bias and Lack of Fairness (A)}: Model's fairness and biases should be tested in both algorithms and data.
\item \textit{Difficulty in reproducibility due to the stochastic nature of parameter initialization (A)}: Many ML-libraries use random initialization of initial states which makes it hard to reproduce issues.
\end{inparaenum} \textit{Standard} category has one challenge: \textit{Lack of standard testing specification (D)}. A model's performance is tested as a whole rather than for a specific input, making it challenging to design a test oracle \cite{barr2014oracle}.

\nd\bf{$\bullet$ SDLC Phase 5. MLSA Deployment} There are four QA challenges in this phase and 3 under \textit{Practice} category: \begin{inparaenum}[(1)]
\item \textit{Insufficient MLOps support into the traditional DevOps process (D)}: There is a lack of end to end deployment pipeline support such as advanced logging, automated rollback, security \cite{paleyes2020challenges}. 
\item \textit{Lack of scalability in real-world scenarios (D)}: Many ML models perform well in a research setting but are not computationally scalable for large scale deployment.
\item \textit{Elevated complexity in ML configuration management (A)}: Rapid experimentation requires to track code, data, parameters, hyperparameters to compare the trade-off of different algorithms, model architecture. 
\end{inparaenum} There is one QA challenges under \textit{Standard} category: 
\textit{Inadequate standard regulation/compliance (D)}. In order to ensure public trust the model needs to be continually checked to determine whether it is following the user data protection acts (HIPPA\cite{annas2003hipaa}, GDPR\cite{voigt2017eu}).

\nd\bf{$\bullet$ SDLC Phase 6. MLSA Maintenance.} The maintenance of MLSA over time is expensive and challenging\cite{sculley_hiddent_debt_2015}. There are 5 QA challenges. \textit{Data} category has two challenges: \begin{inparaenum}[(1)]
\item \textit{Gradual drift and decay of ML models (M)}: ML systems have a predictable performance degradation over time if the models are not updated with new data.
\item \textit{Lack of adversarial robustness due to data dependency (M)}: The deployed ML-application should perform reasonably in the real world and prevent adversarial attacks\cite{schulam2019can, jiang2018trust}.
\end{inparaenum} The \textit{Practice} category has three challenges: \begin{inparaenum}[(1)]
\item \textit{Non-deterministic debugging due to hidden feedback loop (M)}: An ML application learns from data and hence some data-driven feedback could be unseen or remain hidden \cite{sculley_hiddent_debt_2015}. 
\item \textit{Reliability monitoring due to the model's lack of generalizability (M)}: A deployed ML model requires constant monitoring to assess the output against new input and new corner cases. \item \textit{Presence of additional technical debt due to ML adoption into software (A)}: Many undeclared external systems can consume an ML model's output and cause ``visibility debt''~\cite{morgenthaler2012searching}.
\end{inparaenum}

\section{Recommendations and Conclusions} \label{sec:MLSA_opportunities}
We conclude by summarizing the research opportunities to address the MLSA quality assurance challenges we categorized across the six SDLC in \fig\ref{fig:taxonomy_MLSA}. We find that MLSAs and traditional software applications share some common QA challenges like insufficient specifications and privacy. However, as the performance and robustness of MLSAs depend on the quality of the training dataset, the scope and the severity of these QA challenges differ like non-deterministic outcome, implicit biases in data, lack of interpretability, gradual performance decay, hidden feedback loops pose some of the unique challenges for MLSAs, and so on.  To handle ambiguity and lack of standards in \it{MLSA requirement analysis}, research can focus on developing necessary specifications, tools, and metrics for MLSA requirement  analysis  and third-party verification~\cite{vogelsang2019requirements}. For \it{design and planning challenges of MLSA}, research can improve data privacy~\cite{li2020federated, mcmahan2017communication, bonawitz2019towards} and develop specifications and tools~\cite{galhotra2017fairness} to improve model bias and fairness~\cite{mehrabi2019survey}.
For \it{implementation-specific challenges}, research can focus on developing debugging tools~\cite{ma2018mode}, transfer learning~\cite{survey_transfer_learning_2009}, model modularization~\cite{pan2020decomposing}, and interpretable AI~\cite{lipton2018mythos, ribeiro2016should}.
To handle the \it{challenges of testing of MLSA}, more research is needed on bug analysis\cite{zhang_ml_testing_survey_horizons_2020}, regression testing, reinforced learning ~\cite{zhang_ml_testing_survey_horizons_2020}, novel adversarial attacks \cite{dong2018boosting, carlini2017towards, szegedy2013intriguing, mei2015using} and defence techniques~\cite{goodfellow2014explaining, zhang2019defense}.
To address \it{MLSA deployment and maintenance} challenges, future research can focus on developing configuration management and reliability monitoring specification and tools~\cite{schulam2019can, jiang2018trust, saria_safe_reliable_2019}. 





\begin{small}
\bibliographystyle{IEEEbib}
\bibliography{bibliography}
\end{small}

\end{document}